\documentstyle[multicol,aps,amssymb,epsfig]{revtex}

\begin{document}

\draft

\title{Geometrical explanation and scaling of dynamical
heterogeneities in glass forming systems}

\vspace{-0.2cm}

\author{Juan P. Garrahan$^1$ and David Chandler$^2$}

\address{$^1$Theoretical Physics, University of Oxford, 1 Keble Road,
Oxford, OX1 3NP, U.K. \\ $^2$Department of Chemistry, University of
California, Berkeley, California 94720}

\date{March 11, 2002}

\maketitle

\vspace{-0.2cm}

\begin{abstract}
We show how dynamical heterogeneities in glass forming systems emerge
as a consequence of the existence of dynamical constraints, and we
offer an interpretation of the glass transition as an entropy crisis
in trajectory space (space--time) rather than in configuration
space. To illustrate our general ideas, we analyze the one dimensional
($d=1$) Fredrickson--Andersen and East models. Dynamics of such
dynamically constrained systems are shown to be isomorphic to the
statics of $d+1$ dimensional dense mixtures of polydisperse
non--interpenetrating domains. The domains coincide with arrested
regions in trajectory space.
\end{abstract}

\vspace{-0.2cm}


\vspace{-0.5cm}

\begin{multicols}{2}
\narrowtext

A glass forming system, like a supercooled liquid, exhibits a
precipitous onset of slowness. As temperature is decreased in these
systems, typically in a range of a few decades, relaxation times and
viscosities increase by several orders of magnitude, eventually
surpassing experimentally accessible times. For practical purposes,
these systems effectively freeze at the glass transition temperature
$T_{g}$. For reviews see \cite{GlassReviews}. Interestingly, this
dynamical arrest carries no evident static structural signature of
growing length scales. Rather, experiments and simulations show that
supercooled liquids are {\it dynamically} heterogeneous
\cite{DHReviews,Donati99}. Molecules in one region of the liquid
translate or rotate several orders of magnitude faster or slower than
those in a neighboring region. The spatial extent of these dynamical
heterogeneities is mesoscopic, and the time scale of the slowest
domains increases with decreasing temperature at least as fast as the
relaxation time of the system. Such structural behavior seems beyond
description with homogeneous methods like mode coupling \cite{Gotze}
and mean field theories \cite{Kirkpatrick}, and it is widely neglected
in analytical treatments (see, however,
\cite{Franz,Xia,Chamon}). Nevertheless, we show here that for a
broadly applicable mechanism of dynamical arrest, these
heterogeneities are intrinsic to the nature of glass forming systems.

Our central result is that dynamical heterogeneities are a
manifestation of the existence of nontrivial structure in the
trajectories of glassy systems.  This structure associated with
dynamics is independent of any specific static properties. Instead,
the nontrivial dynamical structure is a consequence of local dynamical
rules that significantly restrict the size of accessible trajectory
space. For example, consider a highly compressed (or supercooled)
glass former. Atoms in most regions of space are jammed, making
mobility possible in only a relatively low fraction of spatial
regions. These rare regions are those that are already unjammed, or
those that may be close in space to an unjammed region. In the
evolution of such a system, one therefore expects a clustering of
mobile regions and thus a mesoscopic demixing of mobile and static
regions. Macroscopic demixing is not expected because dynamics should
conserve a canonical distribution. This picture of the origin of
dynamic heterogeneity is in accord with the idea that glassiness is
not necessarily a consequence of either disorder or frustration in the
static interactions but of the existence of effective constraints on
the dynamics of the system \cite{Palmer84}.

The simplest microscopic models that illustrate this view are the
Fredrickson--Andersen (FA) model \cite{Fredrickson84} and the East
model \cite{Jackle91}. They consist of a chain of Ising spins $\sigma
_{i}=\pm 1$ $(i=1\cdots N)$, with trivial Hamiltonian
$H=\sum_{i}\sigma _{i}$, and single spin flip dynamics subject to
local kinetic constraints. In the FA model, a spin can flip if either
of its nearest neighbors is in the up state \cite{Fredrickson84}. In
the East model, a spin can flip only if its nearest neighbor to the
right is up \cite{Jackle91}. The equilibrium behavior of both models
is that of an uncorrelated spin system, with $c=1/\left(
1+e^{1/T}\right) $ being the average concentration of up spins at
temperature $T$. The competition between decreasing the energy and the
need for facilitating spins leads to a glassy slow down at low
temperatures. The relaxation times go as $\tau \propto e^{3/T}$ in the
FA model, and as $\tau \propto e^{1/(T^{2}\ln 2)}$ in the East
model. These models therefore correspond to strong and a fragile glass
formers, respectively. See \cite {Evans} for a short review.

Let us first establish that both the FA and East models display
dynamical heterogeneities. Consider the coarse grained spins
$s_{i}(t;\Delta t ) \equiv (\Delta t)^{-1} \int_{0}^{\Delta t }
dt^{\prime }\sigma_{i}(t+t^{\prime })$. Fast and slow spins in the
time window of width $\Delta t $ at time $t$ will correspond to low
and high values of $s_{i}^{2}(t;\Delta t )$, respectively. The spatial
distribution of $s_{i}^{2}(t;\Delta t )$ will determine the extent to
which the system is dynamically heterogeneous. The heterogeneity made
evident with this field depends on the coarse graining time. For
$\Delta t $ very short, only the trivial uncorrelated static structure
is probed. The same is true for $\Delta t $ much larger than the
relaxation time where ergodicity is restored. For intermediate values
of $\Delta t $, however, we may expect to see spatial structure in the
$s_{i}^{2}(t;\Delta t)$. In Fig.\ 1 we show that this is indeed the
case by plotting the structure factor, $S(k)$, for the field
$s_{i}^{2}(t;\Delta t)$ calculated from simulations of the FA model
(left panel) and East model (right panel). We have used coarse
graining times of about a fifth of the relaxation time, and
simulations were performed using continuous time Monte Carlo
\cite{Newman99} for systems of $10^{5}$ spins averaged over $10^{3}$
samples. For both models, the existence of a correlation length $\xi
_{\Delta t }$ is clear. Moreover, $\xi _{\Delta t }$ grows, albeit
slowly, with decreasing temperature and thus increasing relaxation
time. For the FA model the structure factor decays approximately as
$S(k)\sim k^{-2}$ for large $k$. In the East model, at intermediate
$k$, the structure factor goes as $S(k)\sim k^{-\ln 3/\ln 2}$.  For
larger momenta, the structure factor is oscillatory, a feature which
becomes more pronounced with decreasing temperature.

In order to understand these behaviors, we study the generating
functional for the trajectories of these systems. Consider a
trajectory $\{\vec{\sigma}_{t}\}\equiv (\vec{\sigma}_{0},
\vec{\sigma}_{1}, \vec{\sigma}_{2}, \ldots, \vec{\sigma}_{{\cal T}})$,
where $\vec{\sigma}_{t}\equiv (\sigma _{1t},\sigma _{2t}, \ldots
,\sigma _{Nt})$ denotes the configuration of the $N$ spin system at
time $t$. Each trajectory has a probability $P(\{\vec{\sigma}_{t}\})$,
so that the total probability of going from $\vec{\sigma}_{0}$ to
$\vec{\sigma}_{{\cal T} }$ is the weighted sum of such trajectories,
$Z_{\vec{\sigma}_{0},\vec{\sigma}_{{\cal T} }}\equiv
\sum_{\{\vec{\sigma}_{t}\}}P(\{\vec{\sigma}_{t}\})$. This partition
sum, $Z_{\vec{\sigma}_{0},\vec{\sigma}_{{\cal T} }}$, can be written
as \cite{TBP}
\begin{equation}
Z_{\vec{\sigma}_{0},\vec{\sigma}_{{\cal T} }} =
\sum_{\{\vec{\sigma}_{t}\}} e^{S_{0}[\{\vec{\sigma}_{t}\}]} \left[
\prod_{it}\delta (h_{it})\right] e^{\Delta S[\{\vec{\sigma}_{t}\}]}
\,.
\end{equation}
The first factor in the summand of (1) corresponds to the probability
of a trajectory in the absence of dynamical constraints. Its action is
just that of $N$ non--interacting ferromagnetic Ising chains in
uniform fields, $S_{0}(\{\vec{\sigma}_{t}\}) = \sum_{i=1}^{N}
\sum_{t=0}^{{\cal T} -1} \left( J_{+--} \, \sigma _{it} \sigma _{it+1}
\right.$ $+$ $J_{-+-} \, \sigma _{it} $ $+$ $J_{--+} \, \sigma_{it+1}$
$+$ $\left. J_{+++} \right)$, where $J_{\mu \nu \varepsilon } \equiv
\left\{ \ln [1-\gamma \,\delta t\,(1-c)] \right.$ $+$ $\mu \, \ln
(1-\gamma \, \delta t \, c) $ $+$ $\nu \, \ln [\gamma \,\delta
t\,(1-c)] $ $+$ $\left. \varepsilon \, \ln (\gamma \,\delta
t\,c)\right\} /4$, $\gamma$ is a microscopic rate, and $\delta t$ the
time step. The second factor in the summand of (1) embodies the
kinetic constraint. Only trajectories which satisfy the condition
$h_{it}=0$ at all space--time points are allowed. If $h_{it}$ is
chosen to be a positive number when the constraint is not satisfied,
then $\prod_{it}\delta (h_{it})=\delta ({\cal H})=e^{-\lambda {\cal
H}}$, where ${\cal H}\equiv \sum_{it}h_{it}$ and $\lambda \to \infty
$. The infinite coupling constant means that the space of allowed
trajectories is that of the ground states of ${\cal H}$. The
respective forms of ${\cal H}$ for the FA and East models are
\begin{eqnarray}
{\cal H}_{{\rm FA}} &=&\sum_{it}\left( 1-\sigma _{it}\sigma
_{it+1}\right) \left( 1-\sigma _{i+1t}\right) \left( 1-\sigma
_{i-1t}\right) /8 \, , \nonumber \\ {\cal H}_{{\rm East}}
&=&\sum_{it}\left( 1-\sigma _{it}\sigma _{it+1}\right) \left( 1-\sigma
_{i+1t}\right) /4 \, , \nonumber
\end{eqnarray}
so that the $i$--th term is zero unless the spin $i$ flips and the
facilitating spins are in the down state. The third factor in the
summand of Eq.\ (1) accounts for preservation of norm and detailed
balance. Trajectories which are allowed by the constraints have
different probabilities than they would in the unconstrained case. In
particular,
\begin{eqnarray}
\Delta S_{{\rm FA}} &=&\sum_{it}\left( 1-\sigma _{i+1t}\right) \left(
1-\sigma _{i-1t}\right) \left( \tilde{J}_{-}\sigma
_{it}+\tilde{J}_{+}\right) /4 \, , \nonumber \\ \Delta S_{{\rm East}}
&=&\sum_{it}\left( 1-\sigma _{i+1t}\right) \left( \tilde{J}_{-}\sigma
_{it}+\tilde{J}_{+}\right) /2 \, , \nonumber
\end{eqnarray}
where $\tilde{J}_{\pm }\equiv J_{\pm 00}+\ln [1-\gamma \,\delta t \,
(1-c)\,e^{-\lambda }]/4\pm \ln (1-\gamma \,\delta
t\,c\,e^{-\lambda})/4$. These expressions describe a proper dynamics
for any $\lambda$. When $\lambda=0$, the expressions reduce to the
trivial model of independent spins with unconstrained dynamics. When
$\lambda \to \infty $, they correspond to the FA and East models.

${\cal H}$ and $\Delta S$ introduce competing spatial interactions in
the space of trajectories. Those of ${\cal H}$ are strong, and they
are ferromagnetic in the sense that they favor the clustering of like
spins. In contrast, those of $\Delta S$ are weak and
antiferromagnetic. Moreover, the scaling with distance of interactions
in ${\cal H}$ is different than that for those in $\Delta S$, so that
nontrivial structure in the space of trajectories can be expected. In
Fig.\ 2, we show samples of equilibrium trajectories for the
unconstrained case (top), the FA (bottom left) and East models (bottom
right), at $T=1.0$. The difference between the constrained and
unconstrained dynamics is striking. The trajectories in both the FA
and East models display an extensive number of domains of down
spins. These domains are the origin of the dynamical
heterogeneities. Spins within these domains do not change, so when
coarse grained in time, they correspond to slow regions. To the extent
that lowering temperature and thus energy in an atomic system
coincides with decreasing facilitating spin concentration $c$, the
structures observed in Fig. 2 coincide with the correlation observed
in simulation of supercooled liquids \cite{Donati99}.  In particular,
slow and fast dynamical heterogeneities correlate with regions of low
and high energy, respectively. Since each column of the pictures in
Fig.\ 2 is an equilibrium configuration of the noninteracting $H$, the
structure seen in the trajectories is purely dynamical.

Down spins must form closed domains in a trajectory as a consequence
of the local and causal nature of the dynamical constraints. This fact
is apparent from the illustration in Fig.\ 3.  A spin is able to flip
only if an appropriate neighboring spin is in the up state at the same
time. It therefore will have flipped up previously and/or it will flip
down later. As such, a well defined closed boundary must exist between
regions of up and down spins.
These boundaries are formed out of segments with shapes like those
depicted on the top of Fig.\ 3, with the possibility of different
slopes. In the FA model, the dynamical constraint is spatially
symmetric, and four possible kinds of boundary segments are allowed,
with the restriction that all the segments of the first (second) kind
must be below those of the third (fourth) kind. It follows that spin
down domains must form semi convex polygons like that pictured at the
bottom left of Fig.\ 3. Namely, any spatial line cuts the boundary
only twice. In the case of the East model, only the third and fourth
boundary segments are allowed, which restricts the domains to shapes
like that pictured at the bottom right of Fig.\ 3. Since slow
dynamical heterogeneities correspond to spatial projections of the
compact space--time down spin domains, they are necessarily compact,
while the converse is true for fast regions. This observation is in
agreement with what is found with computer simulations of atomistic
models \cite{Donati99}.

The geometrical construction described above implies that spin down
domains cannot penetrate each other. Therefore, the set of
trajectories maps to the configuration space of a two dimensional
mixture of polydisperse non--interpenetrating objects of all the
possible shapes allowed by the dynamical constraints. This observation
motivates a description of the dynamics in terms of $\rho (l,t)$, the
density of domains of typical spatial size (height) $l$ and extension
in time (length) $t$. The partition function for trajectories is then
$Z\sim \int D\rho (l,t)\exp \left( -\Omega [\rho (l,t)]\right) $,
where $\Omega \left[ \rho \left( l,t\right) \right] $ is the free
energy functional of this density. Whatever estimate is used for
$\Omega \left[ \rho \left( l,t\right) \right] $, it is crucial that it
enforces the constraint
\begin{equation}
\int dt\rho (l,t)=\rho _{{\rm eq}}(l)=c\,e^{-c\,l}.  \label{rhoeq}
\end{equation}
This condition ensures that any spatial cut of the equilibrium
trajectories is an equilibrium configuration. For a density functional
theory, it provides an approximation to the effects of the
antiferromagnetic interactions of $\Delta S$.

In general, $\rho (l,t)=\rho (l|t)p(t)$, where $\rho (l|t)$ is the
probability density of $l$ conditioned on $t$, and $p(t)$ is the
probability density of domains of length $t$. In the case of the FA
model, a domain is bounded by the random walks of two up spins between
successive encounters, and $\rho (l|t)$ can be obtained by standard
arguments \cite{Fisher}
\begin{equation}
\rho (l|t)\sim l^{2}(Dt)^{-3/2}\exp (-l^{2}/Dt)  \label{rho-l-t}
\end{equation}
with $D\sim c$. If the domains were isolated, $p(t)$ would be given by
the probability of first return of a random walker, which goes as
$p(t)\sim t^{-3/2}$ for large $t$, leading to $\langle t\rangle \to
\infty $, and to the formation of unbounded domains. This result,
however, corresponds to only the ferromagnetic part of the
interactions in trajectory space. It is frustrated by the condition
(\ref{rhoeq}). By combining (\ref{rho-l-t}) with (\ref{rhoeq}), we
have instead $p(t)\sim (\tau t)^{-1/2}e^{-\sqrt{t/\tau }}$, with $\tau
\sim D^{-1}c^{-2}\sim e^{3/T}$. This result for $\tau $ is precisely
the relaxation time in the FA model. The forms of $\rho (l|t)$ and
$p(t)$ are verified in simulations of the FA model \cite{TBP}. We also
see that in terms of the scaling variables $l^{*}=c\,l$ and
$t^{*}=t/\tau $, the density $\rho (l^{*},t^{*})$ is independent of
temperature. An illustration of these scaling relations is given in
Fig.\ 4. The typical height of domains grows with the cubic root of
the relaxation time, thus explaining the slow increase of the
correlation length $\xi _{\Delta t }$ with decreasing temperature in
the structure factors of Fig.\ 1. Slow growth and domains being
mesoscopic rather than macroscopic are consistent with experimental
observations \cite {DHReviews}. Moreover, since the trajectories are
extensive in interfaces, the form of the structure factor for large
$k$ corresponds to a spatial projection of Porod's law $S(k)\sim
k^{-(d+1)}$ \cite{Bray}.

In the case of the East model, typical distances and times are related
at low temperatures by a $T$ dependent dynamic exponent $z(T)\sim
1/(T\ln 2)$ \cite{Sollich,Evans}. We therefore expect the conditional
probabilities to be $\rho (l|t)\propto \exp {[-l^{z(T)}/t]}$. The
condition (\ref{rhoeq}) then leads to a stretched exponential form for
the persistence function, $\int_{t}^{\infty }dt^{\prime }p(t^{\prime
})\sim \exp {[-(t/\tau )^{\beta (T)}]}$, where $\beta (T)\sim 1/z(T)$
and $\tau \sim c^{-z(T)}$, in accordance with previous results
\cite{Buhot}. In this case, $\xi _{\Delta t }$ increases even more
slowly with decreasing $T$ than in the FA model. The geometry of the
domains also explains the differences in the FA and East structure
factors shown in Fig.\ 1. In contrast to the FA case, where
neighboring domains can be compressed to the point of contact, the
boundary between domains in the East model cannot be formed just by a
single line of up spins in space-time. Instead, to decrease energy and
thus increase probability, the boundary is wet by smaller domains,
ideally like in Fig.\ 5. This structure has fractal dimension
$d_{f}=\ln 3/\ln 2$, and gives rise to the behavior $S(k)\sim
k^{-d_{f}}$ at intermediate momenta. At larger $k$, $S(k)$ probes the
granular structure of the boundary, and thus exhibits the oscillatory
behavior shown in Fig.1 \cite {Wet}.

The ideas presented here in detail for the FA and East models
generalize to higher dimensions provided the dynamical constraints are
causal and local, so that excitations favor the creation of
neighboring excitations. In general, dynamical constraints will appear
only below a crossover temperature $T_{x}$ or above a corresponding
packing fraction. This can be incorporated into our description by
allowing the coupling $\lambda$ to be a function of temperature or
packing fraction, increasing with decreasing $T$ or increasing
packing. The temperature $T_{x}$ will then coincide with the
so-called ``landscape'' temperature\cite{GlassReviews}.

The domains that thus appear for $T<T_{x}$ and grow with further
decrease in $T$ can lead to a broken symmetry. Consider, for example,
the partition function $Z_{T}(q,{\cal T})$ for trajectories between
configurational fluctuations with overlap $q$ at time difference
${\cal T}$, $Z_{T}(q,{\cal T}) \equiv \sum_{\{ \vec{\sigma}_{0},
\vec{\sigma}_{{\cal T}} \}} \hat{\rho}(\vec{\sigma}_{0}) \,
Z_{\vec{\sigma}_{0},\vec{\sigma}_{{\cal T} }} \, \delta \left(q -
N^{-1} \sum_{i} \delta\sigma_{i0} \delta\sigma_{i {\cal T}} \right)$,
where $\delta\sigma_{it} \equiv \sigma_{it} - \langle \sigma \rangle$,
and $\hat{\rho}$ is Boltzmann distribution. Since $Z_{T}\left( q,{\cal
T} \right)$ is proportional to the number of such trajectories, it is
natural to write it as $Z_T(q,{\cal T}) \propto \exp
[N\,\omega_{T}(q,{\cal T} )]$, where $\omega_{T}(q,{\cal T})$ is the
entropy density in trajectory space. It is approximately the entropy
of mixing of the slow domains in trajectory space. At values of $T$
for which the temporal extension of these domains is much smaller than
${\cal T} $, $\omega_{T}\left( q,{\cal T} \right) $ is extensive in
time, and $Z_{T}(q,{\cal T} )$ is peaked at $q=0$. In this case,
correlation functions are exponential.

Non-zero overlap is probable, i.e., $Z_{T}\left( q,{\cal T} \right) $
is peaked at finite $q$, only when ${\cal T}$ is comparable to or
smaller than the length of typical slow domains. As temperature is
decreased, the size of these domains increases, constricting the
available trajectory space. At a low enough temperature, we may
therefore expect $\omega_{T}\left( q,{\cal T} \right) $ to become
sub-extensive, leading to a probable finite $q$ throughout a
relatively large range of ${\cal T}$. This change in behavior of
$\omega_T\left( q,{\cal T} \right) $ signals a dramatic change in the
number of available trajectories and a corresponding onset of
dynamical arrest. In this picture, therefore, the glass transition
coincides with an entropy crisis in trajectory space, rather than in
configuration space.

This work was supported by the National Science Foundation, the
Glasstone Fund, and Merton College, Oxford.

\vspace{0.4cm}

\begin{figure}[t]
\begin{center}
\epsfig{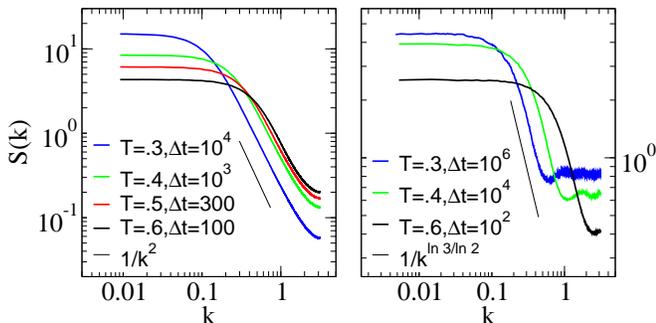}
\caption{Structure factor $S(k)$ of $s_{i}^{2}(t;\Delta t)$ for the FA
(left) and East model (right). The $S(k)$ are the spatial Fourier
transforms of the normalized correlation functions $\langle
s_{i}^{2}(t;\Delta t )s_{j}^{2}(t;\Delta t )\rangle /\langle
s_{i}^{4}(t;\Delta t )\rangle $ ($j=i,i\pm 1,\ldots$).  $\langle \cdot
\rangle$ indicates equilibrium ensemble average, so $S\left( k\right)
$ is independent of $t$.}
\end{center}
\end{figure}

\begin{figure}[t]
\begin{center}
\epsfig{file=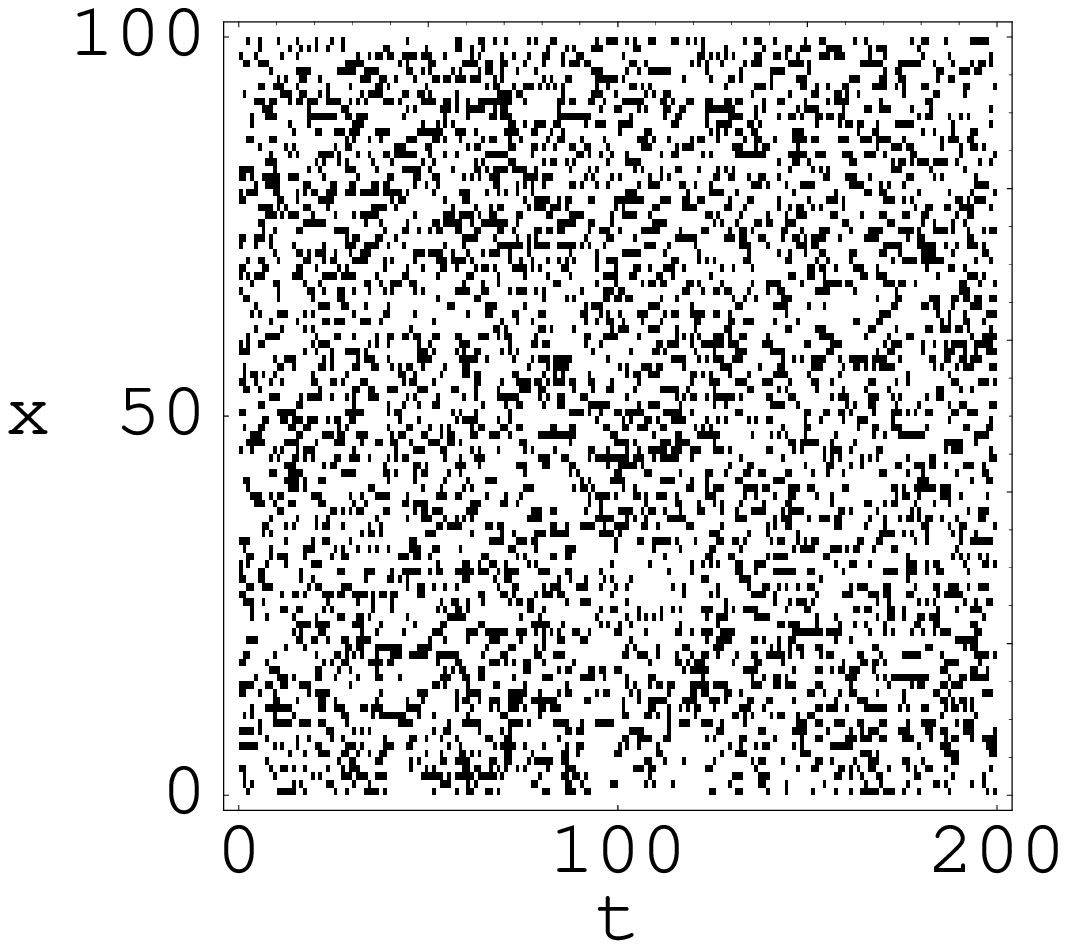, width=1.5in}
\epsfig{file=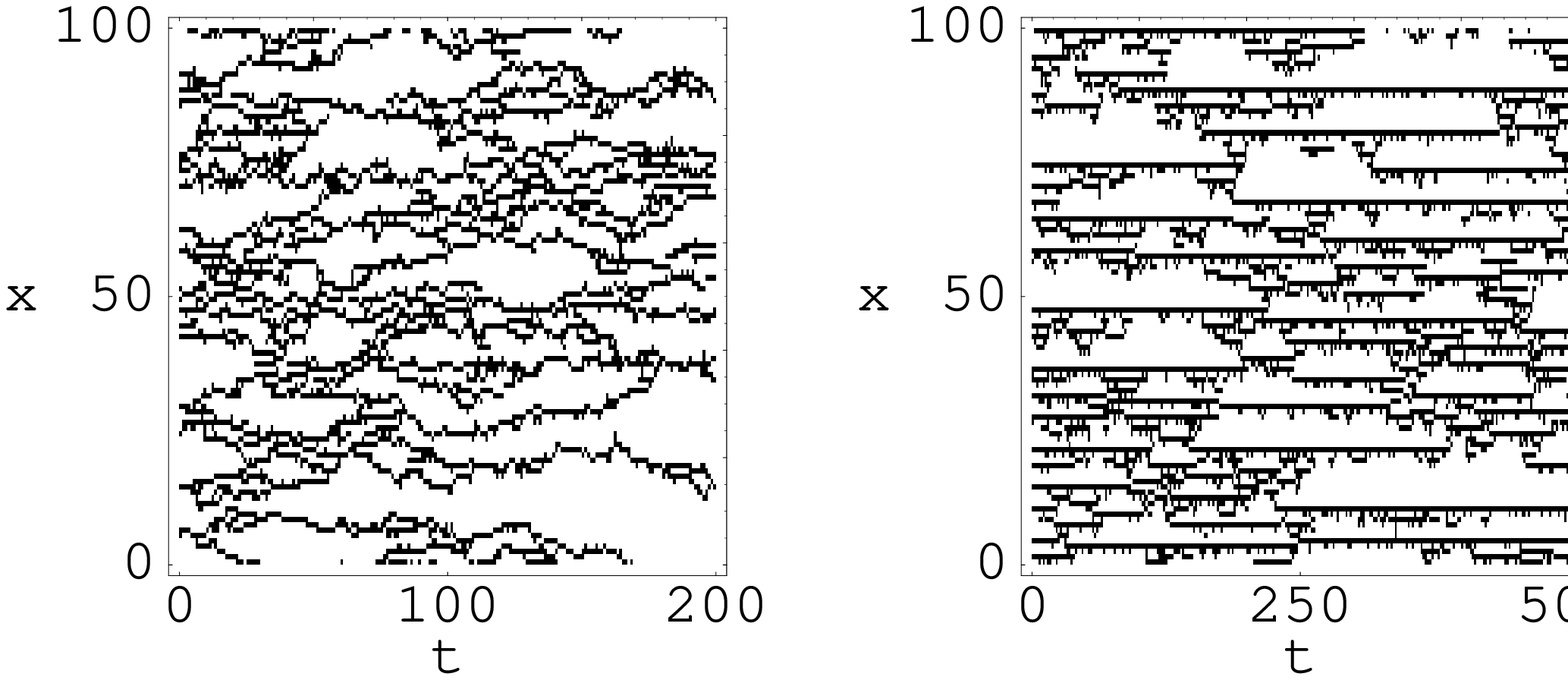, width=3.0in}
\caption{Equilibrium trajectories at $T=1.0$ for the unconstrained
case (top), the FA (bottom left) and East model (bottom right). The
vertical direction is space, corresponding to a spatial window of
systems of size $L=10^5$. The horizontal direction is
time. Black/white correspond to up/down spins. }
\end{center}
\end{figure}

\vspace{-0.8cm}

\begin{figure}[t]
\begin{center}
\epsfig{file=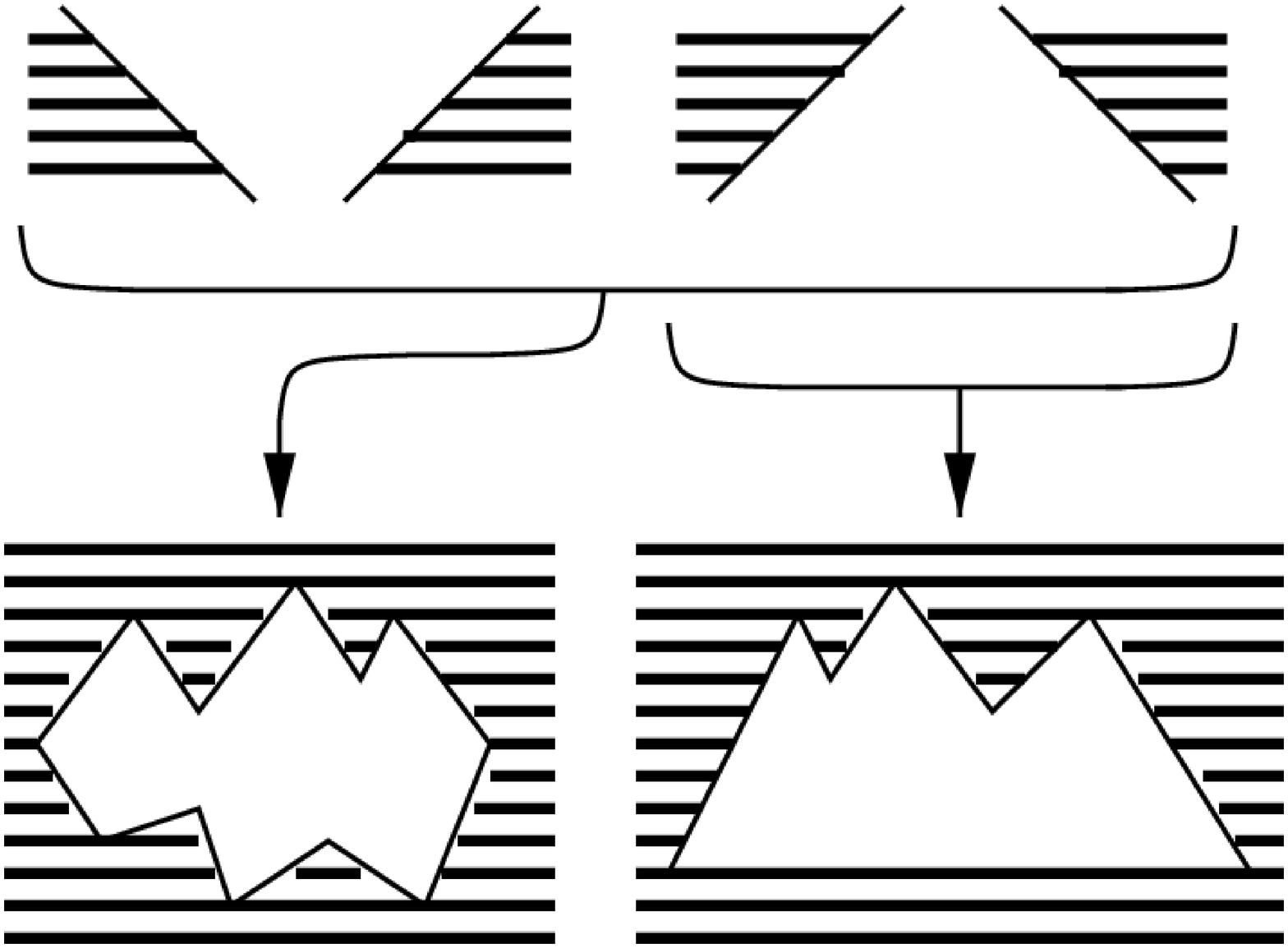, width=1.6in}
\caption{Geometry of slow domains imposed by the dynamical
constraints. Top: allowed boundaries between regions of up (black) and
down spins (white). Bottom: shape of domains in the FA model (left)
and in the East model (right).}
\end{center}
\end{figure}

\vspace{-0.8cm}

\begin{figure}[t]
\begin{center}
\epsfig{file=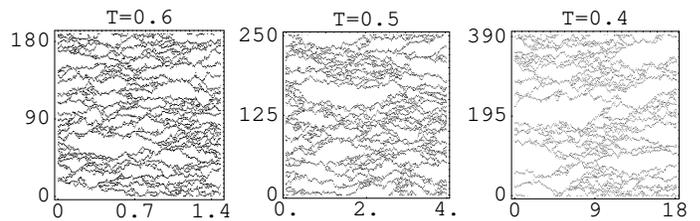, width=3.6in}
\caption{Equilibrium trajectories in the FA model for $T=0.6$, $0.5$,
$0.4$. Vertical direction is space, which scales with $1/c$, and
horizontal is time ($t/10^3$), which scales with $\tau$.}
\end{center}
\end{figure}

\vspace{-0.5cm}

\begin{figure}[t]
\begin{center}
\epsfig{file=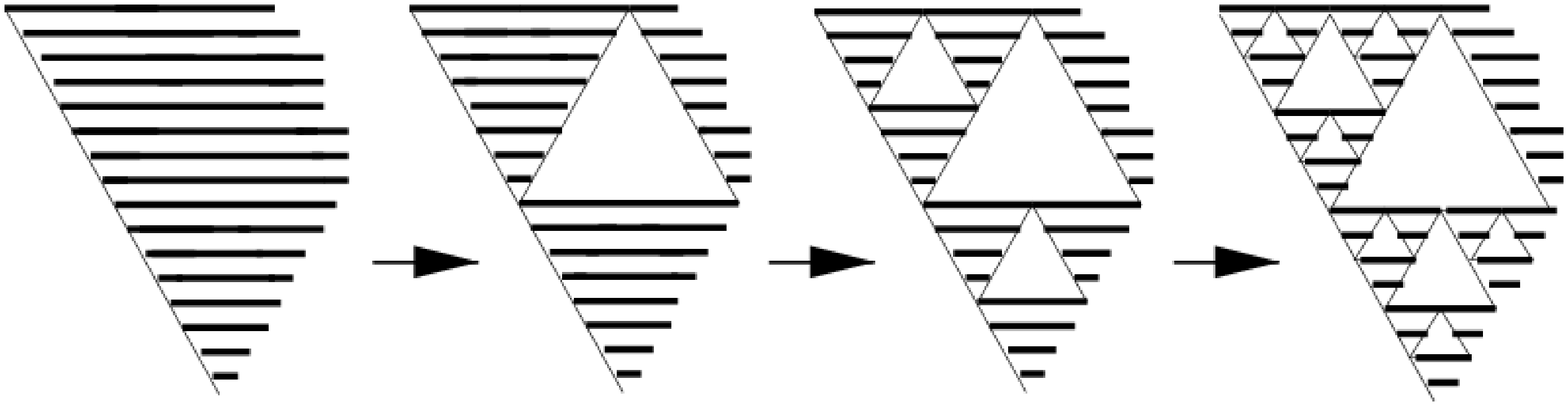, width=2.6in}
\caption{Energetically favored wetting of the boundary of a spin down
domain in the East model. }
\end{center}
\end{figure}

\end{multicols}


\begin{thebibliography}{}
\bibitem{Wet} The idea of wet interfaces in configuration space
appears in a recent equilibrium theory of glass formers \cite{Xia},
but it is difficult to see a connection between that idea pertaining
to statics and the wet interfaces of trajectory space 
discussed here.
\end{thebibliography}
\end{document}